\documentclass[10pt,a4paper]{article}
\usepackage{a4wide}

\usepackage[T1]{fontenc}
\usepackage{amsmath,amsfonts,amssymb,mathbbol}
\usepackage[dvips]{hyperref}
\usepackage{euscript}
\usepackage{graphicx}
\usepackage{psfrag}

\newcommand{\tr}[0]{\mathrm{tr}}
\newcommand{\trm}[0]{\tr_{\mathcal{M}}}  
\newcommand{\trg}[0]{\tr_{R}}  
\newcommand{\trt}[0]{\tr_{\otimes}}      
\newcommand{\str}[0]{\mathcal{S}\trg} 
\newcommand{\detm}[0]{\det_{\mathcal{M}}}  
\newcommand{\detmg}[0]{\det_{\mathcal{M}\otimes R}}  
\newcommand{\detcm}[0]{\det_{\mathbb{C}^2\otimes \mathcal{M}}}  
\newcommand{\detcmg}[0]{\det_{\mathbb{C}^2 \otimes \mathcal{M} \otimes R}} 
\newcommand{\J}{J}

\begin{document}
\date{} 
\title{Non-Abelian generalization of Born-Infeld theory \\ 
  inspired by non-commutative geometry}
\author{Emmanuel Seri\'e$^{\, *, **}$ \and Thierry Masson$^{\, *}$ \and
  Richard Kerner$^{\, **}$} 
\maketitle
\begin{center}
  $^{*}$
  Laboratoire de Physique Th\'eorique (UMR 8627)\\
  Universit\'e Paris XI,\\
  B\^atiment 210, 91405 Orsay Cedex, France\\
\end{center}
\begin{center}
  $^{**}$
  Laboratoire de Physique Th\'eorique des Liquides,\\
  Universit\'e Pierre-et-Marie-Curie - CNRS UMR 7600 \\
  Tour 22, 4-\`eme \'etage, Bo\^{i}te 142, \\
  4, Place Jussieu, 75005 Paris, France\\  
\end{center}

\begin{abstract} 
We present a new non-abelian generalization of the Born-Infeld  Lagrangian. 
It is based on the observation that the basic quantity defining  it
is the generalized volume element, computed as the determinant of a linear
combination of metric and Maxwell tensors. 
We propose to extend the notion of determinant to the tensor product of
space-time and a matrix representation of the gauge group.
We compute  such a Lagrangian explicitly in the case of the $SU(2)$ gauge
group and then explore the properties of static, spherically symmetric
solutions in this model. We have found a one-parameter family of 
finite energy solutions. In the last section, the main properties
of these solutions are displayed and discussed.
\end{abstract}
\vfill
LPT-Orsay 03-51
\newpage

\section{Introduction}
\indent Recently there has been  rising interest in the Born-Infeld nonlinear 
theory of electromagnetism (\cite{born_infeld:34}, \cite{mie:12}) and more general Lagrangians of this type, 
which appear quite naturally in string theories.
Non-Abelian generalizations of a Born-Infeld type Lagrangian were proposed by 
T. Hagiwara in $1981$, (\cite{hagiwara:81}), and more recently, including a
supersymmetric version, by Schaposnik {\it{et al}} (see  (\cite{gonorazky:98},
\cite{christiansen:98}, \cite{grandi:00}) and the references within).
In (\cite{gal'tsov:00}) we analyzed one of the possible 
non-abelian generalizations of the Born-Infeld Lagrangian, and showed the existence
 of sphaleron-like solutions with a qualitative behavior similar to the solutions 
of the combined Einstein-Yang-Mills field equations found by Bartnik and McKinnon
 (\cite{bartnik:88}). 
The non-abelian generalization proposed in (\cite{gal'tsov:00}) was quite straightforward indeed: 
it consisted in the replacement of the electromagnetic field invariants 
$$  F^{\mu \nu} \, F_{\mu \nu} \, \, \, \ \ {\rm and} \, \ \ \, \ \ ^*
F^{\lambda \rho} \, F_{\lambda \rho} = \frac{1}{2}\epsilon^{\mu \nu \lambda \rho} \,
F_{\mu \nu} F_{\lambda \rho} $$
by similar expressions formed by taking  the traces of corresponding Lorentz invariants 
in the Lie algebra space:
$$ g_{ab} \, F^{a \, \mu \nu} \, F^{b}_{\mu \nu} \, \, \, \ \ { \rm and } \, \ 
\ \, \ \ g_{ab} \, ^* F^{a \, \lambda \rho} \, F^{b}_{\lambda \rho} =\frac{1}{2} g_{ab}
\, \epsilon^{\mu \nu \lambda \rho} \, F^a_{\mu \nu} F^{b}_{\lambda \rho} $$
However, except for a straightforward analogy, this expression does not seem to come 
from any more fundamental theory. In addition, this generalization still keeps
a particular dependence on second-order invariants of the field tensor, characteristic for 
a {\it four-dimensional} manifold only; in higher dimensions the determinant would 
lead quite naturally to Lagrangians depending on higher-order invariants of the 
field tensor, too. On the other hand, it is well known that a correct mathematical 
formulation of gauge theories considers the gauge field tensor associated with 
a compact and semi-simple gauge group $G$ as a connection one-form in a principal 
fibre bundle over Minkowskian space-time, with values in ${\cal{A}}_G$, the Lie 
algebra of $G$. In local coordinates we have
\begin{equation}
A = A^a_{\mu}\, dx^{\mu} \, L_a
\end{equation}
where $L_a , \, \ \ a = 1,2,...N=\dim(G)$ is the basis of the adjoint representation 
of ${\cal{A}}_G$. In many cases another representation must be chosen, especially when 
the gauge fields are supposed to interact with spinors (cf. (\cite{domokos:79}, 
\cite{kerner:81}, \cite{kerner:83})). It is always possible to embed the Lie algebra 
in an enveloping associative algebra, and to use the tensor product:
\begin{equation}
A = A^a_{\mu} \, d x^{\mu} \otimes \, T_a \, ,
\end{equation}
where $T_a$ is the basis of the matrix representation of ${\cal{A}}_G$, so that 
now the non-abelian field tensor will have its values in the enveloping algebra:
\begin{equation} F = dA + \frac{1}{2}[A, A] = ( \, F^a_{\mu \nu} \, dx^{\mu}
  \wedge d x^{\nu} \, ) \otimes \, T_a \, .
\end{equation}
Now, in order to reproduce as closely as possible the classical Born-Infeld Lagrangian, 
a natural idea is to embed the space-time metric tensor $g_{\mu \nu}$ also  into the 
enveloping algebra, tensoring it simply with the unit element in the appropriate matrix 
space, i.e. replacing it by $g_{\mu \nu} \otimes {\mathbb 1}_{N}$ ;
then we can add up the metric and the field tensors, and take the determinant in the 
resulting matrix space. This structure of the matrix is similar to the structures found 
in certain realizations of gauge theory in non-commutative matrix geometries 
(\cite{dubois-violette:90}), or in Lagrangians found in matrix 
theories (\cite{dubois-violette:90:II}, \cite{dubois-violette:89:II}). 
Such a Lagrangian has been proposed by Park  (\cite{park:99}) and reads as follows:
\begin{equation}
  S_{Park}[F,g]
  = \int_{\mathbb{R}^4} \alpha \left( \left|\detmg \left( g_{\mu \nu} \otimes 
\mathbb{1}_{d_R}+\beta^{-1}\,F^a_{\mu \nu} \otimes T_a \, \right)\right|^{\frac{1}{2 d_R}} 
 -\sqrt{|g|} \right) \ , 
  \label{park:BI}
\end{equation}
where $\alpha $ and $\beta$ are real positive constants.  
The $2 d_R$-order root is introduced to ensure the invariance of the resulting
action under the diffeomorphisms. 
As a matter of fact, with the root of this order we are able to factorize out the usual 
four-dimensional volume element $\sqrt{|g|} d^4x$ and rewrite the action principle with 
the subsequent scalar quantity:
\begin{align} 
  L_{Park}(F,g) & = \alpha \left(\left| \detmg\left( \mathbb{1}_{4 \times d_R} 
      +  \beta^{-1}\,\hat{F}\,\right)\right|^{\frac{1}{2 d_R}} \, -1\right)
  \label{park:operateur} 
\end{align}  
and,
\begin{align}   
  S_{Park}[F,g] &= \int_{\mathbb{R}^4} L_{Park}(F,g) \sqrt{|g|}d^4x \ ,
\end{align}

where
\begin{align} 
  \hat{F} = \  \frac{1}{2} F^a_{\mu \nu} \hat{M}^{\mu \nu} \otimes T_a \, \ \
  \, \ \ \, \ \ \,
  (\hat{M}^{\mu \nu})^{\rho}_{\sigma}  =  \ g^{\rho \rho'}
  \delta^{\mu \nu}_{\rho' \sigma} \ ,
\end{align}
$\hat{F}$ is an endomorphism of $\mathbb{R}^4 \otimes \mathbb{C}^{d_R}$, and
$M_{\mu \nu}$ denote the generators of the Lorentz group (in the defining
representation). It is also useful to introduce the notation:
\begin{align}  
  \hat{F}^a &= \frac{1}{2} F^a_{\mu \nu} \hat{M}^{\mu \nu} \ .
\end{align}

The generalization of the Born-Infeld (BI) Lagrangian proposed in this paper results in 
a variational principle that leads to a highly nonlinear system of field equations, 
whose general properties can be analyzed using standard techniques (\cite{abrahams:95},
 \cite{lemos:00}, \cite{gibbons:01:II}). Our aim in this article is to check whether 
stationary regular solutions with finite energy can be found as in (\cite{gal'tsov:00}). 
We consider the standard 't Hooft's monopole ansatz which in this particular case leads 
to one ordinary  differential equation for a single function $k(r)$ of radial coordinate $r$. 
The structure of this equation is similar to the one found in (\cite{gal'tsov:00}),
 with a more complicated term corresponding to friction. Nevertheless, the structure 
of solutions and their energy spectrum are very different, as shown in the last sections 
of our article. We have not found solutions joining together two different vacuum 
configurations (called BI {\it sphalerons}), as in (\cite{gal'tsov:00}). We find 
instead a family of solutions labeled by integer winding number $n$, and a real 
parameter bounded from below. The energy integral tends with $n \rightarrow \infty$ 
to the energy of the BI magnetic monopole obtained in (\cite{gal'tsov:00}).

\section{ Non-abelian generalization of Born-Infeld Lagrangian}
\subsection{Basic properties of the abelian case}
\indent
Let us  recall several basic properties of the abelian Born-Infeld Lagrangian, which we would 
like to reproduce in the proposed non-abelian generalization.  In their original paper 
(\cite{born_infeld:34}) Born and Infeld considered the now famous least action principle:
\begin{equation} \label{BI}
  \begin{split}
    S_{BI}[g,F] &= \int_{\mathbb{R}^4} {\cal{L}_{BI}}(g,F) = \int_{\mathbb{R}^4}{L_{BI}(g,F)}
 \sqrt{|g|}d^4x \\               
    &= \int_{\mathbb{R}^4} \beta^2 \left( \sqrt{|\det(g_{\mu \nu}) |} -
      \sqrt{|\det(g_{\mu \nu}+\beta ^{-1} \,
        F_{\mu \nu}\,)\,|} \right) d^4x \\
    &= \int_{\mathbb{R}^4} \beta^2 \left(1-\sqrt{ 1+ \frac{1}{\beta^2} (F,F)
        - \frac{1}{4 \beta^4}(F,{\star F})^2}\right) \sqrt{|g|}d^4x \\
    & = \int_{\mathbb{R}^4} \beta^2\left(1-\sqrt{1+ \frac{1}{\beta^{2}}
        (\vec{B}^2 - \vec{E}^2) - \frac{1}{\beta^{4}}(\vec{E}.\vec{B})^2}\right)
    \sqrt{|g|}d^4x
  \end{split}
\end{equation}
where $d^4x= dx^0 \wedge dx^1 \wedge dx^2 \wedge dx^3$ , $\vec{B}$ is the magnetic field, 
$\vec{E}$ is the electric field, $(F,F)=\frac{1}{2} F_{\mu \nu} F^{\mu \nu}$,  $(F,{\star F})=\frac{1}{4} \epsilon^{\mu \nu \rho \sigma} F_{\mu \nu} F_{\rho \sigma}$ 
and $\epsilon^{\mu \nu \rho \sigma}=\frac{1}{\sqrt{|g|}} \delta^{\mu \nu \rho \sigma}_{0123}$.  

This action can be defined not only on the Minkowskian space-time 
but also on any locally Lorentzian curved manifold, as in the original case. 
 It is useful to recall here three important properties of the Born-Infeld Lagrangian, 
which we want to maintain in the case of the non-abelian generalization, also valid in any finite 
dimension of space-time. 

1) Maxwell's theory (or, respectively, the usual gauge theory with quadratic Lagrangian density) should 
be found in the limit $\beta \to \infty$:
\begin{equation}\label{limite}
  \begin{split}
    {S_{BI}}
    &=  - \int_{\mathbb{R}^4} \frac{1}{2} (F,F) \sqrt{|g|}d^4x + o(\frac{1}{\beta ^2})\\
    &= - \frac{1}{2}\int_{\mathbb{R}^4} F\wedge {\star F} + o(\frac{1}{\beta
      ^2})\\
    &= - \int_{\mathbb{R}^4}\frac{1}{2}(\vec{B}^2 - \vec{E}^2) \sqrt{|g|}d^4x +
    o(\frac{1}{\beta ^2}) \ .
  \end{split}
\end{equation}
2) There exists an upper limit for the electric field intensity, equal to
$\beta$ when there the magnetic component of the field vanishes:
\begin{equation}\label{Mie}
  {L_{BI}}|_{B=0}
  =\beta^2 \left(1-\sqrt{1- \beta^{-2} \vec{E}^2}\right) \ .
\end{equation}
Due to this fact, the energy of a pointlike charge is finite, and the field
remains finite even at the origin. This was the main goal pursued by Mie
(\cite{mie:12}), suggesting the choice of nonlinear generalization of
Maxwell's theory. Indeed, one has for a point charge $e$:
\begin{equation}
  \vec{E}=\frac{e \hat{r}}{\sqrt{e^2+r^4}} \, \ \ \, \ \
  \text{Energy} = \int_0^{\infty} \left(\frac{\sqrt{e^2+r^4}}{r^2} -1 \right
  )r^2 dr  < \infty \ .
\end{equation}
3) The Born-Infeld action principle is invariant under the diffeomorphisms of $\mathbb{R}^4$. 
In this respect, this theory can be viewed as a covariant generalization (in the sense of 
General Relativity) of Mie's theory, as well as an extension of the usual volume element 
$\sqrt{|g|} d^4 x$.

It is also well known that the Born-Infeld electromagnetism has good causality properties 
(no birefringence and no shock waves) as well as interesting dual symmetries 
(electric-magnetic duality, Legendre duality, cf.
\cite{boillat:70} \cite{plebanski:70} \cite{bialynicki-birula}
\cite{gibbons:01:II} \cite{gibbons:95:II}) . Here we shall not consider these aspects
of the theory, our main interest being focused on static solutions.

\subsection{The new non-abelian generalization}

Our starting point is the gauge field tensor associated with a compact and semisimple 
gauge group $G$, defined as a connection one-form in the principal fibre bundle over Minkowskian 
space-time, with its values in ${\cal{A}}_G$,
the Lie algebra of $G$. 
As explained in the Introduction, we chose the representation of the connection in the 
tensorial product of a matrix representation of the Lie algebra   ${\cal{A}}_G$ and the 
Grassmann algebra of forms over $M_4$:
\begin{equation}
  A = A^a_{\mu}\, dx^{\mu} \otimes \, T_a
  \label{connection}
\end{equation}
where $T_a , \, \ \ a= 1,2,...N=\dim(G)$ are anti-hermitian matrices which form a basis of 
the particular representation $R$ of dimension $d_R$ of $\cal{A}_G$, specified later on.

By analogy with the abelian case, we want the Lagrangian to satisfy the following properties: 
\vskip 0.2cm
\indent
1) One should find the usual Yang-Mills theory in the limit $\beta \to \infty$
\vskip 0.2cm
\indent
2) The (non-abelian) analogue of the electric field strength should be bounded from above 
when the magnetic components vanish.
[ To satisfy this particular constraint, we must ensure that the polynomial
expression under the root should start with terms
$1-\beta^{-2}(\vec{E}^a)^2 + ...$ when $\vec{B}^a =0$ ].
\vskip 0.2cm
\indent
3) The action should be invariant under the diffeomorphisms of $\mathbb{R}^4$.
\vskip 0.2cm
\indent
4) The action has to be  real.
\vskip 0.2cm
This enables us to introduce the following generalization of the Born-Infeld Lagrangian density 
for a non-abelian gauge field:
\begin{align}
  {\cal{L}} =\sqrt{g} \,L =  \sqrt{|g|} - \left| \, \det\left( \mathbb{1}_2 \otimes g_{\mu \nu} 
      \otimes \mathbb{1}_{d_R} +\beta ^{-1} \, \J \otimes F_{\mu \nu}^a \otimes T_a \,\right)\,
  \right|^{\frac{1 }{4 d_R}} 
\end{align}
In the expression above, $J$ denotes a $SL(2, \mathbb{C})$ matrix satisfying 
$J^2 = -\mathbb{1}_2$, thus introducing a quasicomplex structure.
This extra doubling of tensor space is necessary in order to ensure that the resulting 
Lagrangian is real. 
We are left with the  root of order $4d_R$, so that the invariance of our action under the 
space-time diffeomorphism is preserved. 
\newline
\indent Let us recall a few arguments in favor of this construction:  
\newline
\indent 
The simplest way to generalize the Born-Infeld action principle to the
non-abelian case seems at first glance the substitution of real numbers by
corresponding hermitian operators, like in quantum mechanics or in
non-commutative geometry.  Then one would arrive at the following expression:
\begin{equation}
  \left\{\begin{array}{lcl}
      U(1)& \leftrightsquigarrow & G \\      
      i F_{\mu \nu}  & \leftrightsquigarrow &  F_{\mu \nu}^a \otimes T_a \\      
      g_{\mu \nu}   & \leftrightsquigarrow & g_{\mu \nu} \otimes
      \mathbb{1}_{d_R} \ ,\\      
    \end{array} \right.
  \label{correspondance}
\end{equation}
where $\mathbb{1}_{d_R}$ and $ iT_a$ are hermitian matrices.  
What remains now to make the generalization complete, is to extend the notion
of the determinant 
taken over the space-time indices in
the usual case.
We propose to replace the determinant of a $4 \times 4$ matrix (denoted hereafter $\detm$) 
by a determinant taken in the tensor product of space-time and
matrix indices of the representation $R$
(denoted hereafter $\detmg$).
Notice that this kind of tensor product of algebras appears in the context of the noncommutative geometry 
of matrices (see (\cite{dubois-violette:90} \cite{dubois-violette:90:II} \cite{dubois-violette:89:II})).
Indeed, the general structure of the connection one-form in these noncommutative geometries 
is very similar to the one in (\ref{connection}).

In this kind of generalization, one would replace the objects in (\ref{BI}) following 
the procedures in  (\ref{correspondance}).
This leads to a complex Lagrangian in the case of a non-abelian structure group. 
Indeed, the determinant $ \detmg ( g_{\mu \nu} \otimes 
\mathbb{1}_{d_R}+\beta^{-1}\,F^a_{\mu \nu} \otimes iT_a \, )$ is not real when  
 $dim({\cal{A}}_G) >1$. 
Therefore we must find a different generalization.

Another possibility would consist of taking anti-hermitian generators tensorized with 
the field $F$.
This was proposed by Hagiwara (\cite{hagiwara:81}) and studied in more detail by Park 
(\cite{park:99}) for the Euclidean case. 
This substitution leads to a Lagrangian satisfying the requirements  1), 3), and  4), 
but not 2) (for details, see  the article by J.-H. Park, (\cite{park:99}) ).

Moreover, Lagrangians obtained with the above choices display invariants of order 
$3$ in the field $F$, destroying the charge conjugation invariance of the theory, 
$F \mapsto -F$, and possibly  leading to indefinite energy densities. 

This is why we propose a third choice. We start from an alternative formulation 
of the abelian version. 
As a matter of fact, one can write the abelian Born-Infeld Lagrangian in the following 
alternative form:
\begin{equation}\label{BI:J}
  S_{BI}[F,g] =\int_{\mathbb{R}^4}  \beta^2 \left( \sqrt{|g|} -
    \left|\detcm \left(\mathbb{1}_2 \otimes g_{\mu \nu} +\beta^{-1} \, \J \otimes
  i F_{\mu \nu} \,\right)\,\right|^{\frac{1}{4}} \right) d^4x \ ,
\end{equation}
where $\J$ is a $2 \times 2 $ complex matrix whose square is equal to $-\mathbb{1}_2$.
 The Lagrangian is independent of the choice of $\J$ as can be easily seen. 
In  (\ref{BI:J}) (see also (\ref{correspondance})), the imaginary unit $i$ 
can be considered as the anti-hermitian generator of $\mathfrak{u}(1)$.
In our formula (\ref{BI:J}), we use an obvious notation for the space on which the determinant 
is defined.
With the correspondence displayed in (\ref{correspondance}), we end up with the following 
action principle:
\begin{align}\label{notreBI}
  S[F,g] &=\int_{\mathbb{R}^4} \alpha \left( \sqrt{|g|} - \left| \,
  \detcmg\left( \mathbb{1}_2 \otimes g_{\mu \nu} \otimes \mathbb{1}_{d_R}
  +\beta ^{-1} \, \J \otimes F_{\mu \nu}^a \otimes T_a \,\right)\,\right|^{\frac{1 }{4 d_R}} 
\right) d^4x  \ ,
\end{align}
satisfying all the requirements we asked for, 1), 2), 3), and 4), by taking $\J$ in $SL(2, \mathbb{C})$. The Lagrangian is again independent of the choice of $\J$.
In particular, we find the usual abelian Lagrangian if we replace ${T_a}$ by ${i}$ 
and set $d_R=1$.
\\
It was supposed in (\ref{notreBI}) that  $\alpha$ and $\beta$ are real positive constants. 
It is clear that only the root of degree $4 d_R$ will lead to an expression where 
$\sqrt{g}$ can be factorized out as an overall factor.
This enables one to rewrite the action using a purely scalar quantity as follows:
\begin{align} 
  L(g,F) &= \alpha \left( 1 - \left|\detcmg\left(\mathbb{1}_2 \otimes
  \mathbb{1}_{4 \times d_R} +\beta
  ^{-1} \, \J \otimes \hat{F} \right) \,\right|^{\frac{1}{4 d_R}}\, \right)  ,  
  \label{notreBI:2}
\end{align}
so that
\begin{align} 
  S[g,F] &= \int_{\mathbb{R}^4} L(g,F) \sqrt{|g|}d^4x \ ,
\end{align}
where $\hat{F}=\frac{1}{2} F_{\mu \nu}^{a} \hat{M}^{\mu \nu} \otimes T_a $ as defined
in the introductory section.
 
\section{ Explicit computation of the determinant }
\subsection{General remarks}
The determinant defined in (\ref{notreBI:2}) can be written in several
equivalent forms:
\begin{subequations}\label{determinants}
  \begin{align}
    & \detcmg\left(\mathbb{1}_2 \otimes \mathbb{1} + \beta^{-1} \, 
\J \otimes \hat{F}\right)\label{det1}\\
    &= \detcmg\left(  s \otimes \mathbb{1} + \beta^{-1}\, 
 s \J \otimes \hat{F}\right) \label{det2}\\
    &= \detmg\left(\mathbb{1} + \beta^{-2} \, \hat{F}^2\right)\label{det3} \ ,
 \end{align}
\end{subequations}
where $s$ and $\J$  are elements of $SL(2, \mathbb{C})$, $\J$ satisfying $J^2=-\mathbb{1}$.
For example, choosing  $s=i\sigma_2$
and $ s \J = -i\sigma_3$  in  (\ref{det2}), we get the following determinant:
\begin{equation}\label{matrice_de_commutation}
  \begin{vmatrix}
    -i \beta^{-1} \hat{F} & \mathbb{1} \\
    -\mathbb{1} & i \beta^{-1} \hat{F}
  \end{vmatrix} = |g|^{-2 d_{R}} \begin{vmatrix}
   -i \beta^{-1}  F^a_{\mu \nu} \otimes T_a & g_{\mu \nu} \otimes \mathbb{1} \\
    - g_{\mu \nu} \otimes \mathbb{1} & i  \beta^{-1}  F^a_{\mu \nu} \otimes T_ a
  \end{vmatrix} \,
\end{equation}
which is a straightforward generalization of the determinant considered 
by Schuller  (\cite{schuller:02}). Following Schuller's idea, the matrix
~(\ref{matrice_de_commutation}) in the abelian case can be interpreted as
the matrix defining  commutation relations between the coordinates in the
phase space of a relativistic point particle minimally coupled to the
Born-Infeld field. Similarly, we can extend this interpretation to the case
of coordinates taking their values in an appropriate Lie algebra, i.e., by imposing the following relations:
\begin{equation}
  \begin{split}
    [ X_\mu,X_\nu ] &=  - \frac{1}{e \beta^2} F_{\mu \nu}^a \otimes T_a\\
    [ X_\mu,P_\nu ] &= -i g_{\mu \nu} \otimes \mathbb{1}\\
    [ P_\mu,P_\nu ] &= e F_{\mu \nu}^a \otimes T_a \ ,
  \end{split}
\end{equation}
with
\begin{equation}  
    X_\mu := X^a_\mu \otimes -i T_a \, , \ \ \, \ \ \, \ \ 
    P_\mu := P^a_\mu \otimes -i T_a \, .
\end{equation}
On the other hand, the particular form (\ref{det3}) enables us to check that the Lagrangian is
indeed real, and at the same time it represents an obvious generalization
of the abelian Born-Infeld action in the form given in (\cite{schuller:02},
 and the references therein). 
It is worthwhile to note that if one
chooses  $\J = -i \sigma_3$ in (\ref{det1}) the determinant can
be written as an absolute value of a complex number. 
Indeed, one has
\begin{equation}
  \begin{vmatrix}
    \mathbb{1} -i \beta^{-1} \hat{F} & 0 \\
    0 & \mathbb{1}+ i \beta^{-1} \hat{F}
  \end{vmatrix}
  =
  |\detmg\left(\mathbb{1} -i  \beta^{-1} \hat{F}\right)|^2 \label{module} \ .
\end{equation}
We shall use this particular form of the determinant in the subsequent
computations. \\

\subsection{Comparison with the symmetric trace prescription}
Let us recall a useful formula relating the determinant of a linear
operator $M$ to traces: 
\begin{equation}\label{trace}
  \begin{split}
    \left(\det(1+M)\right)^\beta 
    & = \exp\left({\beta \,\tr(\, \log(1+M)\,)}\right)\\
    & = \sum_{n=0}^{\infty} \ \sum_{\substack{ \underline{\alpha}=(\alpha_1, \cdots ,
 \alpha_n) \\   \in [S_n]}} (-1)^n \prod_{p=1}^{n} \frac{1}{\alpha_p !} 
\left(- \frac{\beta \, \tr(M^p)}{p} \right)^{\alpha_p} \ ,
  \end{split}
\end{equation}
where $\underline{\alpha} \in [S_n]$ and  $[S_n]$ is the set of equivalence
classes of the permutation group of order $n$. The multi-index
$\underline{\alpha}$ is given by a Ferrer-Young  diagram or equivalently satisfies the relation,
\begin{align}
  \sum_{p=1}^{n}\,  p \, \alpha_p & = n  \ , \  \alpha_p \geqslant 0 \ .
\end{align}
 
Using this trace formula, we can develop our Lagrangian up to any 
order in $F$. In order to avoid ambiguities, we shall denote by $ \trm$  
the trace taken over the space-time indices, by $ \trg$ the trace over the representation 
indices, and by $\trt$ the trace over the tensor product of these two spaces. 
For the sake of simplicity, we have absorbed the scale factor $\beta^{-1}$ in the definition 
of the field tensor $F$. 
When needed, the appropriate powers of $\beta^{-1}$ can easily be recovered. 
Following (\ref{det3}), we have,
\begin{equation}\label{traces}
  \begin{split}
    & \left( \detmg\left(1+\hat{F}^2\right) \right)^{\frac{1}{4d_R}}\\
    & =\sum_{n =0}^{\infty} \sum_{\underline{\alpha}=(\alpha_1, \cdots ,
      \alpha_n)} (-1)^{n}
    \prod_{k=1}^{n} \frac{1}{\alpha_k !} 
    \left(\frac{- \trt (\hat{F}^{2k})}{4 d_R \times k} \right)^{\alpha_k}  \\
    & =\sum_{n =0}^{\infty} \sum_{\underline{\alpha}=(\alpha_1, \cdots ,
      \alpha_n)} (-1)^{n} \prod_{k=1}^{n} \frac{1}{\alpha_k !}
    \prod_{\substack{m=1 \\ \alpha_k \neq 0}}^{\alpha_k} \left( - \frac{
      \trm(\hat{F}^{a^m_1} \cdots\hat{F}^{a^m_{2 k }} )}{4 k} \times
    \frac{\trg( T_{a^m_1} \cdots T_{a^m_{2 k}})}{d_R} \right) \ ,
  \end{split}
\end{equation}
where $ \underline{\alpha} \in [S_n] $ satisfies $  \sum_{k=1}^{n}k\alpha_k = n $ . \\
We can compare the resulting expansion with the symmetrized trace prescription
given by Tseytlin in (\cite{tseytlin:97}). With the notation adopted above, we have

\begin{multline}
  \frac{1}{d_R}\str  \left( \detm\left(1+i\hat{F}^a T_a\right) \right)^{\frac{1}{2}} = 
  \frac{1}{d_R}\str  \left( \detm\left(1+\hat{F}^a\hat{F}^b T_a T_b\right) \right)^{\frac{1}{4}}\\
  \shoveleft{ \quad    = \frac{1}{d_R}\str \sum_{n =0}^{\infty} \sum_{
      \underline{\alpha}=(\alpha_1, \cdots , \alpha_{n})} (-1)^n \prod_{ k=1
      }^{n} \frac{1}{\alpha_k !}  \left(- \frac{ \trm(\hat{F}^{a_1}
        \cdots\hat{F}^{a_{2 k}} )}{4 k} T_{a_1}
      \cdots T_{ a_{2 k} } \right)^{\alpha_k} }\\
  \shoveleft{ \quad   =\sum_{n =0}^{\infty} \sum_{ \underline{\alpha}=(\alpha_1, \cdots ,
      \alpha_{n})} (-1)^n \Biggl( \prod_{k=1}^{n} \frac{1}{\alpha_k !}
    \prod_{\substack{m=1 \\ \alpha_k \neq 0}}^{\alpha_k} \left(-
      \frac{\trm(\hat{F}^{a^m_1} \cdots\hat{F}^{a^m_{2 k}}) }{4 k } \right)
    \times }\\
  \times   \frac{1}{d_R} \str \left( \prod_{k=1}^{n} \prod_{m=1}^{ \alpha_k}
    T_{a^m_1} \cdots T_{a^m_{2 k}} \right) \Biggr) \,
\label{Str:2}
\end{multline}
Now we can easily compare the series resulting from similar expansions
of two different Lagrangians: the symmetrized trace prescription, and the
generalized determinant prescription, i.e. comparing (\ref{traces}) and
(\ref{Str:2}) with the corresponding expansion of the abelian version
of Born-Infeld electrodynamics. In both cases, the third-order and higher 
odd-order invariants that are possible in a non-abelian case, do not appear 
(as they are absent in the abelian version, of course) .   

Let us compare, up to the fourth order, the expansion in powers of $F$ of 
the two Lagrangians. 
Our Lagrangian (\ref{notreBI}) yields the following series:
\begin{multline}
  L[F,g] \simeq -\frac{1}{4{d_R}} \trt\hat{F}^2 + \frac{1}{8 {d_R}}
  \trt\hat{F}^4 -\frac{1}{32 {d_R}^2}(\trt\hat{F}^2)^2\\
  \shoveleft{ \qquad \quad     \simeq -\frac{1}{2} (F^a,F^b)K_{ab} +\frac{1}{8}(F^a,F^b)(F^c,F^{d}) (-
    K_{ab}K_{cd} +K_{abcd}+K_{acbd})} \\ 
  +\frac{1}{8} (F^a,{\star
    F}^b)(F^c,{\star F}^d)K_{acbd} \ ,
\end{multline}
whereas the symmetrized trace prescription of (\cite{tseytlin:97}) gives:
\begin{equation}
  \begin{split}
    L_{Sym}[F,g] & = \frac{1}{{d_R}}\str(\mathbb{1} - \sqrt{\detm(\mathbb{1} +
      i \hat{F})})\\
    & \simeq \frac{1}{d_R} \str ( -\frac{1}{4} \trm \hat{F}^2 + \frac{1}{8}
    \trm {F}^4 -
    \frac{1}{32} (\trm \hat{F}^2)^2 )\\
    & \simeq - \frac{1}{2} (F^a,F^b) K_{ab} + \frac{1}{8} \left((F^a,
    F^b)(F^c,F^d)+(F^a,{\star F}^b)(F^c,{\star F}^d)\right) K_{\{abcd\}} \ ,
  \end{split} 
\end{equation}
with $ K_{\{abcd\}} =\frac{1}{3} (K_{ab}K_{cd}+K_{ac}K_{bd}+K_{ad}K_{bc} +
\frac{1}{4} S^e_{ab} S_{cde} + \frac{1}{4} S^e_{ac} S_{bde} + \frac{1}{4}
S^e_{ad} S_{bce})$.
As usual we note
\begin{equation}
  T_a T_b  =- g_{ab} \mathbb{1} + \frac{1}{2} C_{ab}^{c}
  T_{c}+ \frac{i}{2} S_{ab}^{c} T_{c} \ ,
\end{equation}   
where  $g_{ab}=\frac{c_R}{d_R} \delta_{ab}$ , $S_{cab}=g_{cd} S_{ab}^{d}$ is
completely symmetric and real,
$C_{cab}=g_{cd} C_{ab}^{d}$ is completely antisymmetric and real, and,
\begin{align}
  K_{a_1 \cdots \ a_n} = \frac{(-1)^{[\frac{n}{2}]}}{d_R}\trg(T_{a_1} \cdots T_{a_n}) \ .
\end{align}

\subsection{Explicit calculus for G=$SU(2)$}

\newcommand{\Tr}[1]{t_{#1}}
\newcommand{\Trf}[1]{\frac{1}{2}\trt \left(\frac{\hat{F}^{#1}}{#1}\right)} 
\newcommand{\Trff}[2]{\frac{1}{#2 !}\left(\frac{1}{2} \trt \left(\frac{\hat{F}^{#1}}{#1}\right) \right)^{#2}}

We use the fundamental representation of  $G=SU(2)$, with generators
defined by $T_a=-\frac{i}{2}\sigma_{a}$. In order to simplify the
calculus, we have rescaled the formula (\ref{module}) replacing $\beta$ by
$1/2$, so that it compensates the factor $1/2$ in the definition of $T_a$. 
It is useful to note that 
in the formula (\ref{module}), the expression $\detmg(\mathbb{1} -i \beta^{-1} \hat{F})$ 
is a perfect square (as noticed already in (\cite{park:99})). As a matter
of fact, one can multiply this determinant by $ 1 =\detmg ( \mathbb{1} \otimes -i\sigma_{2}) $, 
to obtain
\begin{align}
  \detmg\left(\mathbb{1} - 2 i \hat{F}\right) &= \detmg \left( \mathbb{1} \otimes (-i
  \sigma_2) +
  \hat{F}^a \otimes (i\sigma_2 \sigma_a)\right) \\
  &= |g|^{-2} \detmg \left( g_{\mu \nu} \otimes (-i\sigma_2) + F^a_{\mu \nu}
  \otimes (i\sigma_2 \sigma_a)\right) \ .
\end{align}

It is easily seen that the matrix in the last expression
is antisymmetric, so its determinant is a perfect square. This implies that the highest power
of $F$ in the expansion of $\exp(\frac{1}{2} \tr \log(1+ 2 i \hat{F}))$ 
is $4 $ ; therefore 
\begin{align}
  \detmg ( \mathbb{1} + 2 i \hat{F})
  &= \left( \exp\left({\frac{1}{2} \tr \log ( 1 + i \hat{F} )}\right) \right)^2\\
  & = \left( 1 + \Trf{2} - i \Trf{3} - \Trf{4} +
  \Trff{2}{2} \right)^2  \\
  &= \left( 1 + \frac{\Tr{2}}{4} - i \frac{\Tr{3}}{6} - \frac{\Tr{4}}{8}
  +\frac{\Tr{2}^2}{32} \right)^2 \ ,
\end{align}
where $\Tr{i}= \trt \left( \left(\hat{F} \right)^i\right)$.\\

Using formula (\ref{module}), we get
\begin{equation}
  \label{bina:su2}
  L= 1-\sqrt[4]{(1+2P-Q^2)^2 + (2K_3)^2} \ ,
\end{equation}
where
\begin{equation}
  \left\{\begin{array}{lclcl}
      2P  & =& \frac{1}{4} \Tr{2} &=& (F^a,F_a)\\
      Q^2 & = &\frac{1}{8}\Tr{4}-\frac{1}{32}\Tr{2}^2 &=& \frac{1}{4}(F^a,{\star F}^b)
      (F^c,{\star F}^d) K_{acbd}\\
      K_3 & =& - \frac{1}{12}\Tr{3} &=& \frac{1}{6}\epsilon_{abc} \trm(\hat{F}^a \hat{F}^b 
\hat{F}^c)
    \end{array} \right. \ .
\end{equation}
It is also interesting to note that our Lagrangian depends exclusively on 
three invariants of $F$, (the third-order invariant entering via its square),
although the determinant can lead to the expressions up to the eighth order
in $F$. In this particular case, there exist many relations between the
traces, so that the complicated expressions can finally be simplified and
expressed as functions of three invariants only, even though there are eight for a
general $SU(2)$ Lagrangian(cf (\cite{roskies:77}, \cite{anandan:78})).

\section{Spherically symmetric static configurations}

\subsection{The magnetic ansatz and equations of motion}

Our aim now is to study static, spherically symmetric solutions of purely
``magnetic'' type. They are given by the so-called 't Hooft-Polyakov ansatz 
(\cite{'thooft:74}):
\begin{equation}
  \begin{split}
    A & = \frac{1-k(r)}{2 } UdU^{-1} \ \ \text{with} \  U = e^{i \pi T_r} \\
    &=  (1-k(r)) [ T_r, dT_r] =  (1-k(r)) \  (T_{\theta} \sin\theta d\varphi - 
T_{\varphi} d\theta) \\
    &= \frac{1-k(r)}{ r^2} \  ( \vec{r} \wedge  \vec{T} ) \cdot \vec{dx} \ ,
  \end{split}
\end{equation}
where the usual notation is used.
When expressed in components, the same formula becomes: 
\begin{equation}
A^a_k = \frac{(1 - k(r))\,}{r^2} \  \epsilon^a{}_{km} \, x^{m} \, ,
\end{equation}
where
$$a,b,c... = 1,2,3 \, ; \, \ \ i,j,k...= 1,2,3 \, ; \, \ \ \epsilon^{a}{}_{km} =
\epsilon^{aij} \, g_{ik} \, g_{jm} \, .$$
The notion of spherical symmetry for gauge potentials in Yang-Mills
theory has been analyzed by P. Forgacs and N.S. Manton in (\cite{forgacs:80}); see also (\cite{bertrand:92}).
The most general form for a spherically symmetric $SU(2)$  gauge potential is often
called ``the Witten ansatz'' (cf \cite{witten:77}); an exhaustive discussion of 
its properties can be found in (\cite{volkov:98}). When this form of potential
is chosen, there remaines a residual  $U(1)$ symmetry preserving the field, and the situation
can be interpreted as an abelian gauge theory on two dimensional de~Sitter space, 
coupled to a complex scalar field $w$ with a Higgs-like 
potential. Then the problem is parametrized by four real functions  $a_0$, $a_1$, 
$Re(w)$, and $Im(w)$ (we use the notation introduced
in (\cite{volkov:98})). Fixing the gauge enables one to set $a_1=0$. Next,
one can eliminate $a_0$ if one restrain the solutions to the
``magnetic'' type only. In the static case, the remaining equations of motion possesses
a first integral (due to the residual global $U(1)$ symmetry).
The condition that the energy must be finite at infinity forces it to vanish
in this case. This means that we can choose the phase
of the function $w$ at will, thus reducing the form of the  potential to the one
proposed by 't~Hooft in 1974 (\cite{'thooft:74}).  

Therefore, the only nonvanishing components of the curvature $F$ can be identified
as the  ``magnetic'' components of the Yang-Mills field:
\begin{align}
  B^a_i &=\frac{1}{e r^2} \left[  \hat{r}_i \hat{r}^a (1-k^2) - r  k' \, P^a_i \, \right] \ ,
\end{align}
where  $\hat{r}_i = \frac{x_i}{r}$ and,  $P^a_i=\delta^a_i - \hat{r}^a \hat{r}_i$
is the projection operator onto the subspace perpendicular to the radial direction.

The only nonvanishing invariants of the field appearing in the Lagrangian density
can now be expressed by means of the spherical variable $r$, one unknown real
function $k(r)$, and its first derivative $k'(r)$:
\begin{equation}
  \begin{split}
    2P &= \frac{1}{ r^4} \left[ (1-k^2)^2 + 2 (r k')^2 \right]\\
    K_3 &=\frac{1}{ r^6} \left( (1-k^2) (r k')^2 \right)\\
    Q^2  &=  0
  \end{split}
\end{equation}
Then the action takes on the following form:
\begin{equation}
  S =\int \, \left( 1 - \left\{\left( 1 + \frac{(
      1-k^2)^2 + 2 ( r k')^2 }{r ^4} \right)^2 + \frac{4}{r^{12}} (1- k^2)^2 (r k')^4
  \right\}^{1/4} \  \right) r^2  d r \ .
\end{equation}
For the subsequent analysis, it is very useful to change the independent
variable by introducing its logarithm $\tau = \log(r)$. Then the action can be
expressed as follows: 
\begin{equation}
  S =\int   (1 - \sqrt[4]{A} ) \, e^{3 \tau} \, d \tau \ ,
\end{equation}
where
\begin{displaymath}
  \left\{ 
    \begin{array}{ll}
      A & =(1+a^2+2b)^2+4a^2 b^2 = (1+a^2)((1+2b)^2+a^2)\\
      a & =(1-k^2) /r^2\\
      b &=\dot{k}^2 /r^4
    \end{array} \right.
\end{displaymath}
Now the equation of motion can be written as:
\begin{equation}
  A_{k} + A_{\dot{k}} ( \, \frac{3}{4} \frac{\dot{A}}{A}-3 \, ) - \frac{d}{d
  \tau} A_{\dot{k}} = 0 \, ,
\end{equation}
or equivalently, in a more standard form:
\begin{equation}
  \left\{ \begin{array}{ll}
      \dot{k} &= u \\
      \dot{u} &= \gamma(k,u,\tau) u + k (k^2-1) 
    \end{array} \right.
\label{systeme_dynamique}
\end{equation}
with
\begin{equation}
  \begin{split}
    \gamma (k,u,\tau) &= 1 - 2 \  \frac{ u^2 + 2u k (1- k^2) + (1- k^2 )^2}{ r^4 + (1-
      k^2)^2} \\
    & + \ \frac{
      6 u (1- k^2)\left[ k u^2 + 2u (1-k^2) + k ( 1- k^2)^2 \right] 
      \left[ r^4 + 2 u^2 + (1- k^2)^2\right]}{\left[ r^4 + (1-
      k^2)^2 \right]\left[ (r^4+2 u^2)^2 + (1-k^2)^2(r^4+6 u^2) \right]} \ .
  \end{split}
\end{equation}
The coefficient $\gamma$, which plays the role of dynamic friction, is quite
similar to the one found in  (\cite{gal'tsov:00}) (except for a missing
factor 2, due to a printing error). In the usual Yang-Mills theory with
the same ansatz, the corresponding factor is just $\gamma_{YM} = 1$.

The system (\ref{systeme_dynamique}) is not autonomous (i.e., some of the
coefficients depend explicitly on the variable $\tau$), so that the qualitative
analysis of solutions should be performed in an extended three-dimensional phase 
space $(\tau,k,u)$ (see for example (\cite{chernavsky:78})). Of course, one
cannot expect to find true singular points, because the ``time'' variable $\tau$
never stands still. Instead, one can find asymptotic behaviors of function $k$
whose dominant terms for $\tau \rightarrow - \infty$ ($r \rightarrow 0$)
or for $\tau \rightarrow \infty$ ($r \rightarrow \infty$) satisfy the equations
of motion up to a required order, neglecting infinitely small terms. However,
for $r \rightarrow \infty$ there are two genuine fixed points $(k=1,u = 0)$ and 
$(k=-1,u = 0)$.
Having found these asymptotic expansions, we then try to extend them from both 
sides so that they can meet and produce a regular solution valid for all values 
of $\tau$.

Although our equations display asymptotic expansions analogous to those found
in (\cite{donets:97}, \cite{dyadichev:00}, \cite{gal'tsov:00}), 
careful analysis shows that solutions of the Bartnik-McKinnon
type  (\cite{bartnik:88}) are excluded here.

\subsection{Asymptotic expansions}

\indent
We have found two expansions in positive powers of $r$ which satisfy the equations 
of motion up to a certain finite order in $r$ near  $r=0$. The first one depends
on two free parameters $k_0$ and  $a$, and starts with the following expressions:
\begin{align}\label{dev:r=0:1}
  k= k_0+ a r - k_0 \left(\frac{ 5 a^2 }{6 g}+ \frac{g}{12
    a^2}\right) r^2 + \frac{a^8 (52-70 g) - 9 a^4 g^3 + (g-1) g^4}{108 a^5 g^2} r^3+
  O(r^4) \ ,
\end{align}
where $g=1-k_0^2$,  $a \neq 0$ and $ g \neq 0$.
This expansion displays a certain similarity with the expressions found in 
(\cite{donets:97}, \cite{dyadichev:00}), which depend on 
the same parameter $k_0$.

The second one depends on only one free parameter $b$, and starts as follows:
\begin{align}\label{dev:r=0:2}
  k=\pm \left(  1- b r^2 + \frac{3 b^2 + 92 b^4 + 608 b^6}{10+200 b^2 + 1600
  b^4}\,r^4 + O(r^6) \right)
\end{align}

Near  $r=\infty$, the Taylor expansion can be made with respect to
$r^{-1}$.  It depends on  one free parameter, denoted by  $c$:
\begin{align}\label{dev:inf}
  k=\pm \left( 1- \frac{c}{r} + \frac{3 c^2}{4 r^2} + O(\frac{1}{r^3}) \right) \ .
\end{align}
It is remarkable that the asymptotic behavior at $r = \infty$ is 
the same here (up to the order $O(r^{-7}$)) as the corresponding behavior of 
the spherically symmetric static ansatz in the usual Yang-Mills theory, 
which makes it easier to interpret the characteristic integrals as magnetic 
charge, energy, etc.

Taking these expansions as the  first approximation either at $r=0$ or
at $r = \infty$, we then use standard techniques in order to generate
solutions valid everywhere. It is interesting to note that, when we 
started from infinity, no fine-tuning was necessary, and an arbitrarily
fixed constant $c$ would lead to a solution which, when extrapolated
to $r=0$, would define a particular pair of values of constants
$k_0$ and $a$. On the contrary, starting from $r=0$, arbitrarily
chosen values of $k_0$ and $a$ would not necessarily lead to good
extrapolation at $r= \infty$. We shall discuss the properties of  numerical
solutions so obtained in the following subsection.
 
\subsection{Numerical solutions}

The search for numerical solutions was based on the same method as in 
(\cite{donets:97}, \cite{dyadichev:00}, \cite{gal'tsov:00}).
With the  expansions  (\ref{dev:r=0:1}) and  (\ref{dev:inf}), we evaluate
the initial conditions used as starting point for the
numerical integration of equation (\ref{systeme_dynamique}).

The three parameters occurring in the asymptotic expansions (two at $r=0$
and one at $r= \infty$) are interrelated by two constraint equalities, therefore
the solutions can be labeled by only one real parameter.
We chose to index the solutions with the parameter $c$ of (\ref{dev:inf}), with $c>0$,
or its logarithm $\tau_c=\log(c)$.

\begin{figure}[!h]
\psfrag{n1}[c][c]{\small$n=1$}
\psfrag{n2}[c][c]{\small$n=2$}
\psfrag{n3}[c][c]{\small$n=3$}
\psfrag{n4}[c][c]{\small$n=4$}
\psfrag{k}[c][c]{$k$}
\psfrag{t}[c][c]{$\tau$}
\psfrag{u}[c][c]{$u$}
\includegraphics[width=0.9\textwidth]{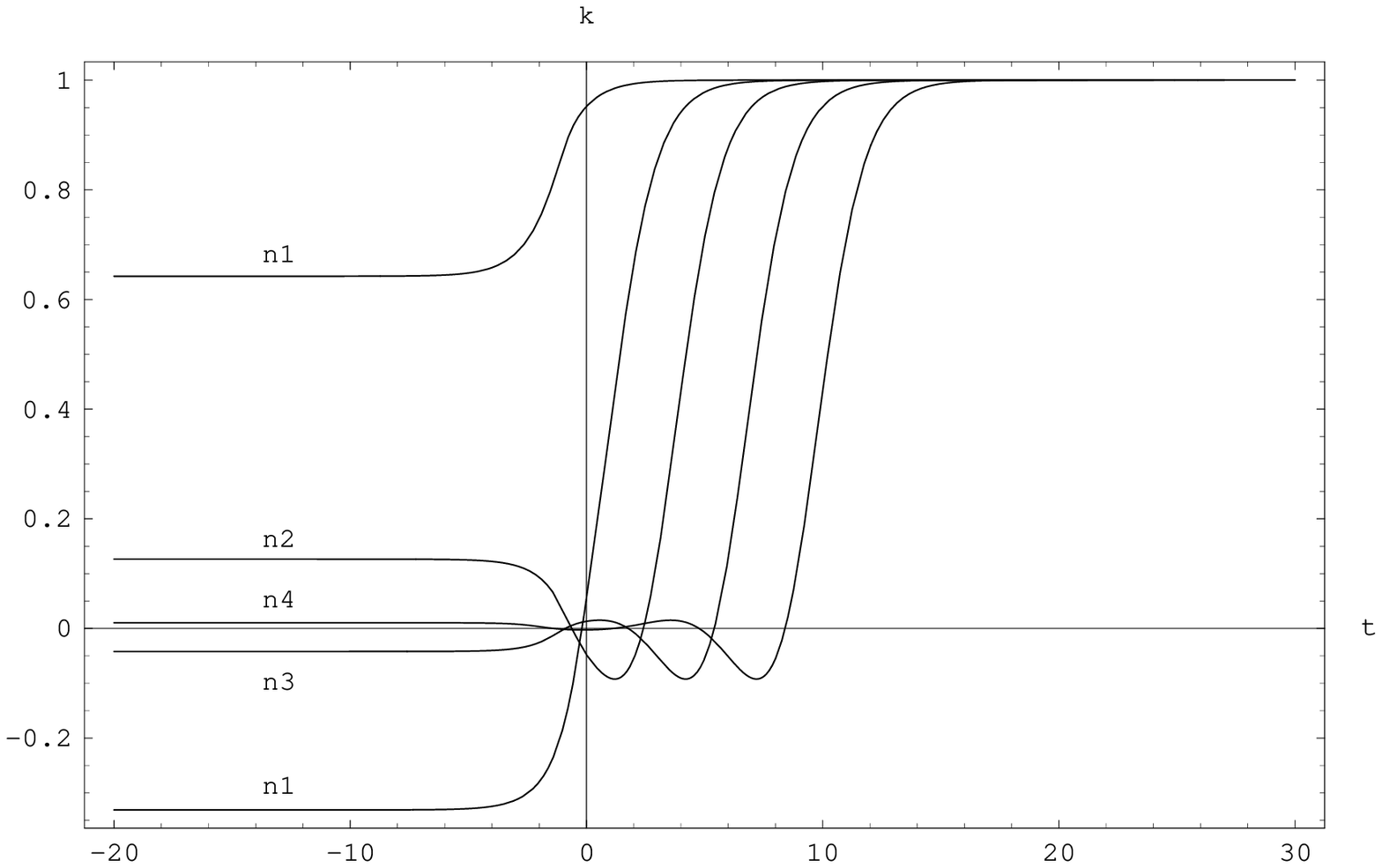}
\includegraphics[width=0.9\textwidth]{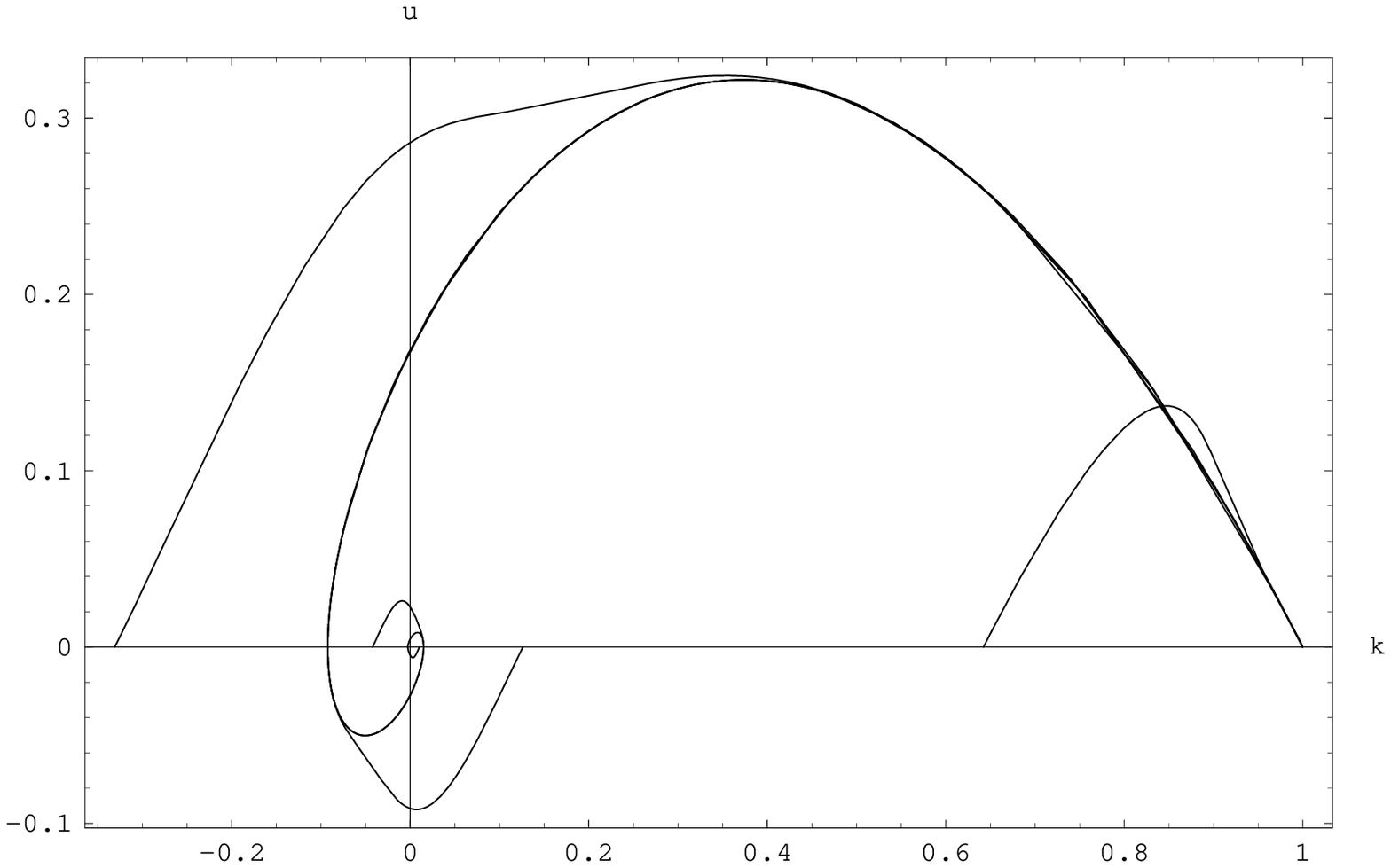}
\caption{Plots of solutions for the parameters $\tau_c=-3, 1.2, 4, 7, 10$.}
\label{some_plots}
\end{figure}

As in the Bartnik-McKinnon case, we can assign to each solution an integer 
$n$, with $n-1$ denoting  the number of zeros of the function  $u$ or the
winding number of the corresponding trajectory in the phase plane  $(k,u)$,
as seen in Fig.~\ref{some_plots}, where a few solutions are plotted.
When the parameter $\tau_c$ goes from $- \infty$ to $+ \infty$, we observe that
this integer $n$ grows from $1$ to $\infty$. 
At certain special values of the parameter $\tau_c$, this integer increases by $1$.
Here are the first critical values of $\tau_c$: 
\begin{center}
  \begin{tabular}[]{|c|c|c|c|c|c|c|c|}
    \hline
    $\tau_c$ & 1.658 & 4.781 & 7.510 & 10.092 &  13.218 &  16.530 & 19.813 \\   
    \hline
  \end{tabular}
\end{center}
The two graphs in Fig.~\ref{some_plots} should  be combined in order to give a correct representation
of the trajectories as they appear in the extended three-dimensional phase space
including the variable $\tau = \log r$. 
The graph on the left represents the cut
$k, \tau$, and the graph on the right represent the cut $k,u$, i.e., the usual phase space of
the function $k(r)$ and its first derivative $u = \dot{k}$. 
One can see some trajectories on the plane $k,u$ with various winding numbers.

Our solutions do not interpolate between the two singular points at $k=1$ and
$k= -1$, but between the singular point at $k=1$ for $r = \infty$ and a certain
value $k_0$ (related to $\tau_c$) which is always lower than $1$ and bigger 
than $-1$ (as a matter of fact $k_0=0$ is a solution). 
This is radically different from the sphaleronlike solutions
or solutions of Bartnik-McKinnon type found in (\cite{bartnik:88}, \cite{gal'tsov:00}).

\begin{figure}[!h]
  \psfrag{n1}[c][c]{\small$n=1$}
  \psfrag{n2}[c][c]{\small$ n=2$}
  \psfrag{n3}[c][c]{\small$n=3$}
  \psfrag{n4}[c][c]{\small$n=4$}
  \psfrag{n5}[c][c]{\small$n=5$}
  \psfrag{n6}[c][c]{\small$n=6$}
  \psfrag{n7}[c][c]{\small$n=7$}
  \psfrag{E}[c][c]{$E$}
  \psfrag{tc}[c][c]{$\tau_c$}
  \includegraphics[width=0.9\textwidth]{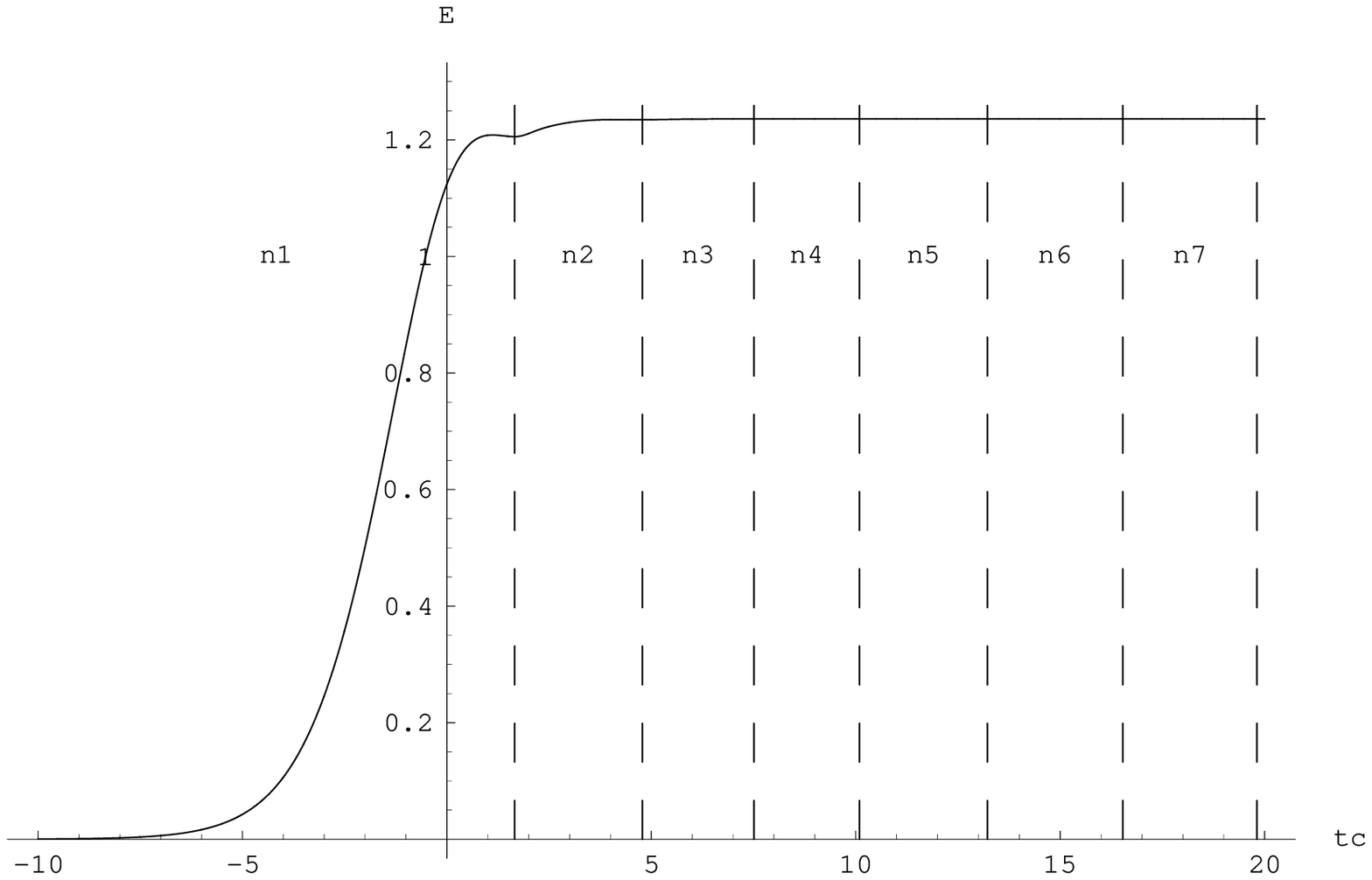}
  \includegraphics[width=0.4\textwidth]{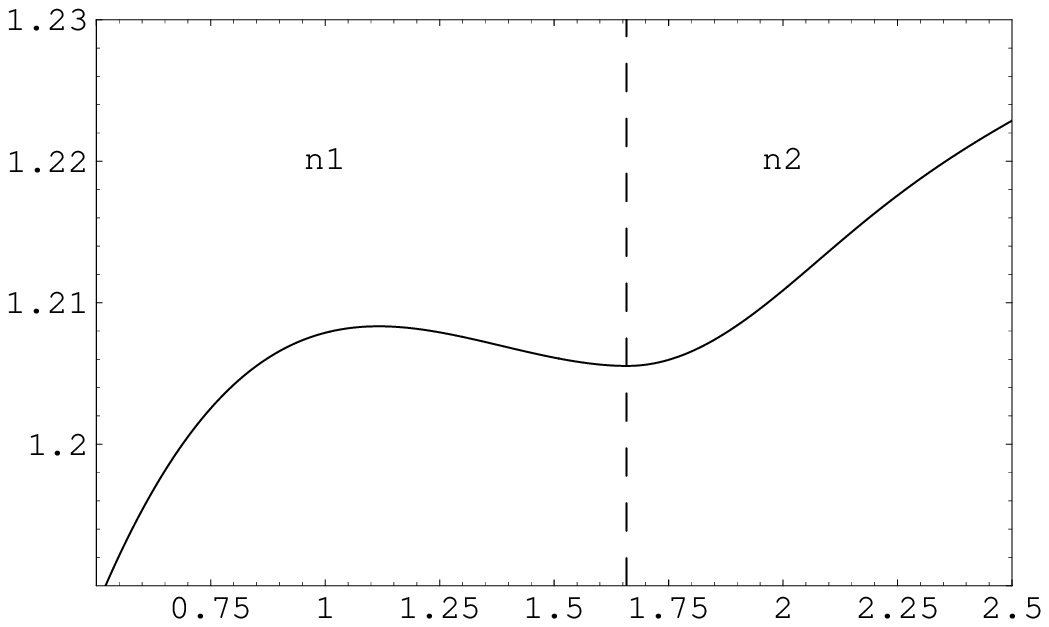}
  \hspace{.5cm}
  \includegraphics[width=0.4\textwidth]{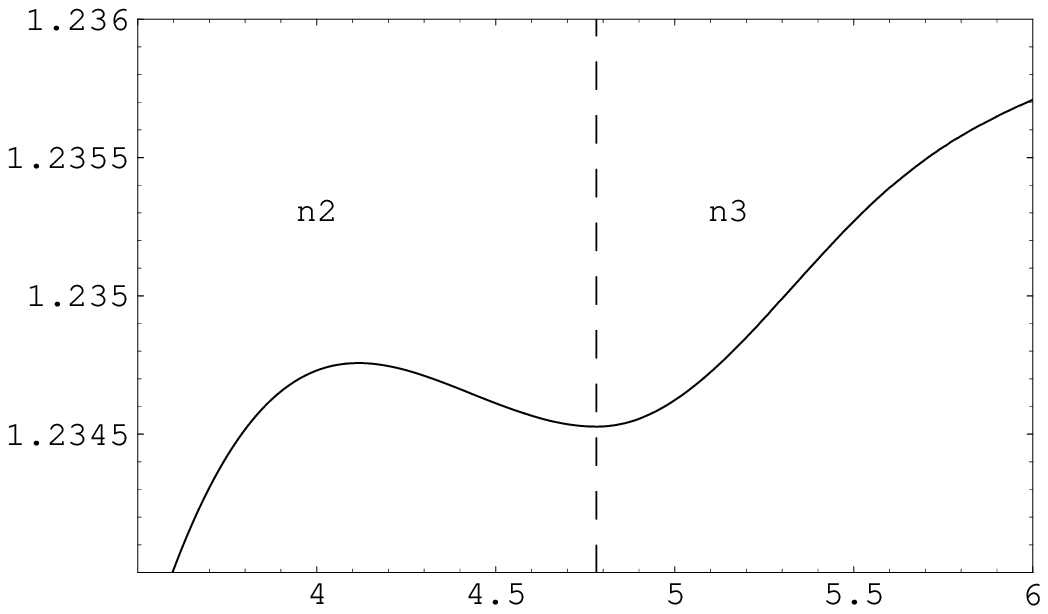}
  \caption{Energy as function of the parameter $ \tau_c$, with local minima visible 
(the magnification is 100 times higher for the second minimum display).}
  \label{fig:E}
\end{figure}

The two parameters  $k_0$ and $a$ of (\ref{dev:r=0:1}) are functions of the
parameter $\tau_c$. We have evaluated the energy  $E$ of the solutions and the values 
of the parameter $k_0$ for $\tau_c$ varying from $-10$ to $20$. 
The energy $E$ is represented as a function of the parameter $\tau_c$ in Fig.~\ref{fig:E}.
This figure represents two enlargements of the upper graph with the precision of $10^{-2}$
in order to show local minima of the energy curve.
The energy minima of each class of solutions are found near
the critical values of the parameter $\tau_c$, and as far as we can judge, given 
the precision of
numerical calculus employed,  coincide with their positions on the $\tau_c$-axis. 
Supposing that the solutions attaining local minima of energy are  stable, we
conjecture that these most stable solutions can be grouped in couples, with
winding numbers $n$ and $n+1$, starting with the couple $n=1, n=2$. 
The energies converge to the limit $E_{\tau_c=\infty}= E_{n=\infty}  = 1.23605...$, 
which coincides with the energy of the pointlike magnetic Born-Infeld monopole 
computed in (\cite{gal'tsov:00}).

\begin{figure}[!h]
\psfrag{n1}[c][c]{\small$n=1$}
\psfrag{n2}[c][c]{\small$n=2$}
\psfrag{n3}[c][c]{\small$n=3$}
\psfrag{n4}[c][c]{\small$n=4$}
\psfrag{n5}[c][c]{\small$n=5$}
\psfrag{n6}[c][c]{\small$n=6$}
\psfrag{n7}[c][c]{\small$n=7$}
\psfrag{ko}[c][c]{$k_0$}
\psfrag{tc}[c][c]{$\tau_c$}
\psfrag{ddko}[c][c]{$\displaystyle\frac{d^2 k_0}{d \tau_c^2}$}
\includegraphics[width=0.9\textwidth]{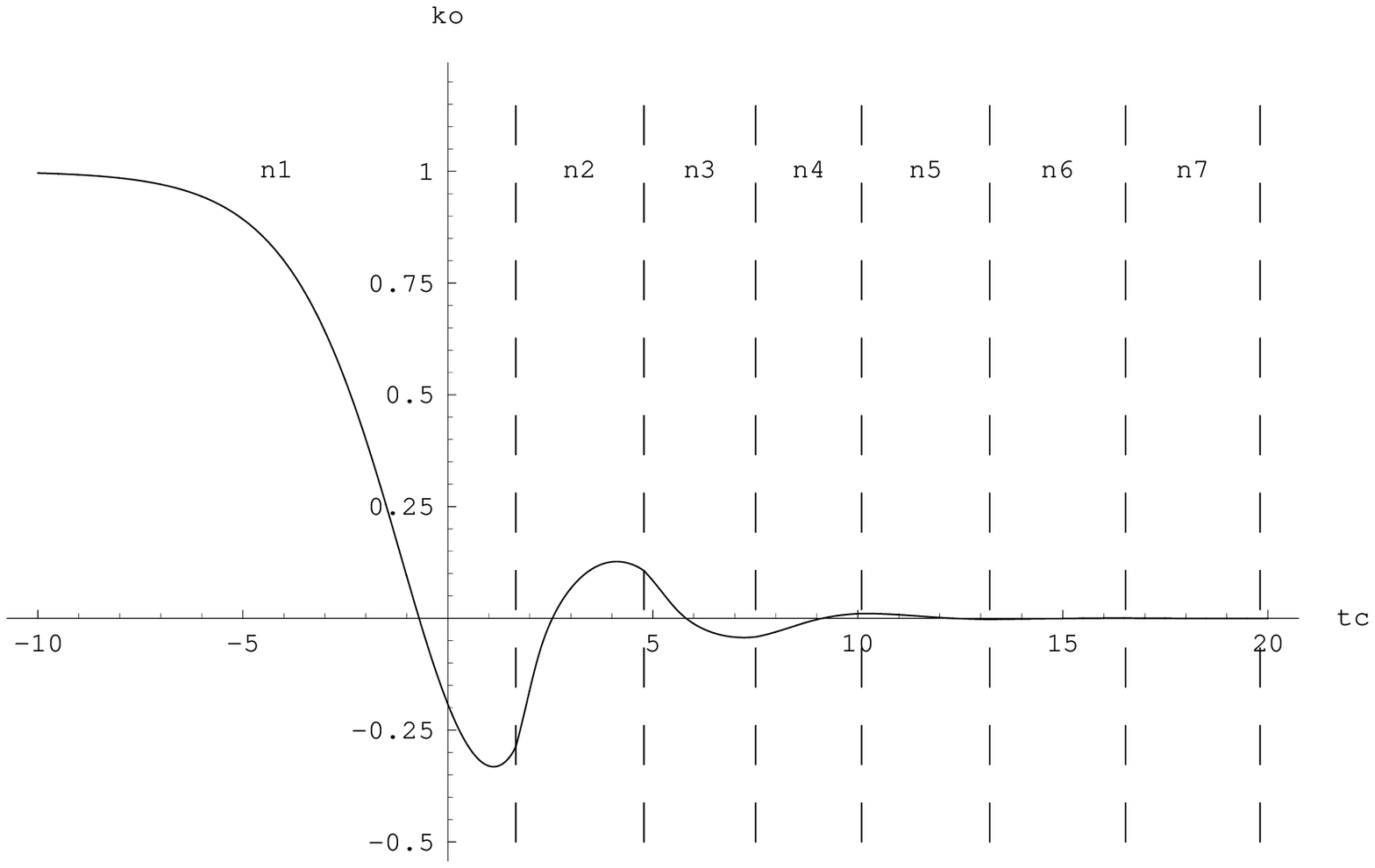}\\ 
\rule{0pt}{1cm}\\
\includegraphics[width=0.9\textwidth]{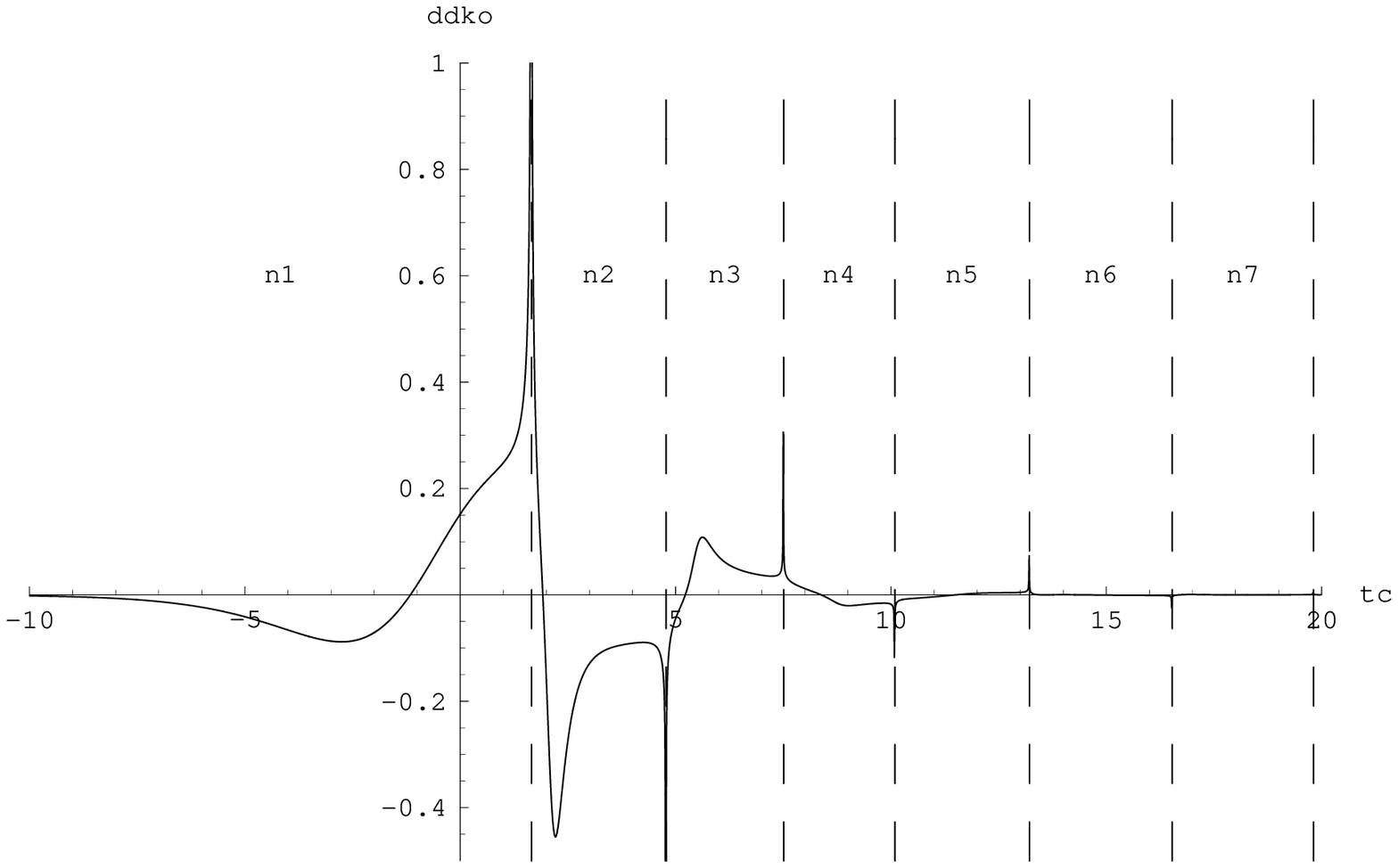}
\caption{$k_0$, function of $\tau_c$, and its second derivative. The singularities of
second derivative  $d^2k_0/d\tau_c^2$ occur at values of  $\tau_c$ which coincide
with the change of winding number $n$ . }
\label{fig:ko}
\end{figure}

The last two graphs in Fig.~\ref{fig:ko} show the specific features of the dependence of the 
parameter $k_0$ (the initial value of function $k$ at $r=0$) with respect to $\tau_c$. 
The dependence is smooth only between the critical values of parameter $\tau_c$, 
at which the change of winding
number $n$ occurs, as can be viewed in the second graph where the second derivative 
of $k_0$ with respect to $\tau_c$ is plotted.

It is important  to notice that our version of generalized non-abelian
Born-Infeld theory is quite different from the symmetrized trace prescription.
Nevertheless, the nonpolynomial character of the Lagrangian, common to all
generalizations, still ensures a very rich spectrum of solutions, although
very different and specific to the choice of the Lagrangian. All our solutions
tend to the genuine vacuum configuration at $r \rightarrow \infty$, but their
behavior near the origin $r = 0$ is very different from the sphaleronlike
solutions. At the origin, our solutions look like monopole configurations
whose magnetic charge has been renormalized, as suggested in
(\cite{donets:97}), where the constant $1- k_0^2$ is also integrated in this manner.
\bigskip 

\section*{Acknowledgments}
We wish to express our thanks to M. Dubois-Violette, D.V. Gal'tsov,
Y. Georgelin and C. Schmit for many enlightening comments.

\clearpage
\bibliographystyle{utphys}

\bibliography{biblio_articles}

\end{document}